\documentclass[journal=jctcce, manuscript=article]{achemso}

\usepackage{chemformula} 
\usepackage[T1]{fontenc} 
 \usepackage[colorlinks=true,linkcolor=blue,urlcolor=blue,citecolor=blue]{hyperref}
\usepackage{array}
\usepackage{lipsum}
\usepackage{bm}
\usepackage{subfigure}
\usepackage{amssymb}
\usepackage{multirow}
\usepackage{tabularx}
\usepackage{amsmath}
\usepackage{braket}
\usepackage{caption}
\usepackage{float}
	
\usepackage{ulem}

\SectionNumbersOn

\author{Tobias Dornheim}
\email{t.dornheim@hzdr.de}
\affiliation{Center for Advanced Systems Understanding (CASUS), Helmholtz-Zentrum Dresden-Rossendorf (HZDR), D-02826 G\"orlitz, Germany}

\author{Sebastian Schwalbe}
\affiliation{Center for Advanced Systems Understanding (CASUS), Helmholtz-Zentrum Dresden-Rossendorf (HZDR), D-02826 G\"orlitz, Germany}

\author{Zhandos A. Moldabekov}
\affiliation{Center for Advanced Systems Understanding (CASUS), Helmholtz-Zentrum Dresden-Rossendorf (HZDR), D-02826 G\"orlitz, Germany}

\author{Jan Vorberger}
\affiliation{Insitute of Radiation Physics, Helmholtz-Zentrum Dresden-Rossendorf (HZDR), D-01328 Dresden, Germany}

\author{Panagiotis Tolias}
\affiliation{Space and Plasma Physics, Royal Institute of Technology (KTH), Stockholm, SE-100 44, Sweden}

\title{\textit{Ab initio} path integral Monte Carlo simulations of the uniform electron gas on large length scales}

\abbreviations{IR,NMR,UV}

\begin{document}







\begin{abstract}
The accurate description of non-ideal quantum many-body systems is of prime importance for a host of applications within physics, quantum chemistry, material science, and related disciplines. At finite temperatures, the gold standard is given by \textit{ab initio} path integral Monte Carlo (PIMC) simulations, which do not require any empirical input, but exhibit an exponential increase in the required compute time for fermionic systems with increasing the system size $N$. Very recently, it has been suggested to compute fermionic properties without this bottleneck based on PIMC simulations of fictitious identical particles. In the present work, we use this technique to carry out very large ($N\leq1000$) PIMC simulations of the warm dense electron gas and demonstrate that it is capable of providing a highly accurate description of investigated properties, i.e., the static structure factor, the static density response function, and local field correction, over the entire range of length scales.
\end{abstract}

\begin{figure}[H]
    \centering
    \includegraphics[width=0.5\textwidth]{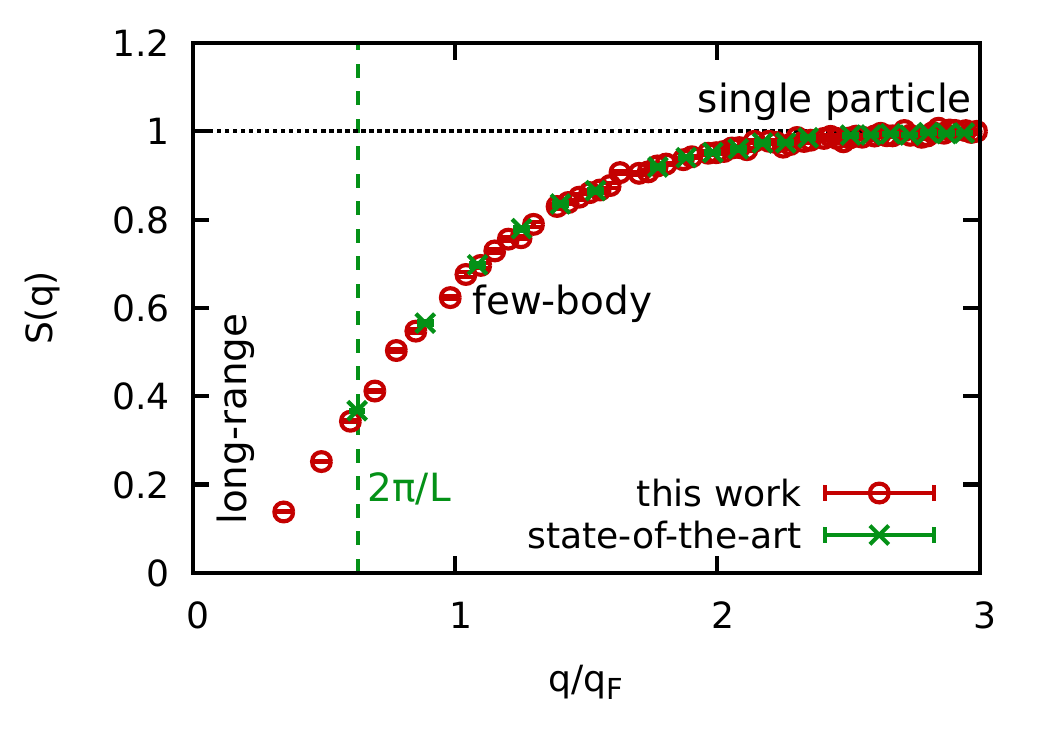}
    \caption*{\textbf{Table of Contents (TOC)/Abstract graphic.}}
    \label{fig:my_label}
\end{figure}


The accurate description of fermionic quantum many-body systems is a paramount task within physics, quantum chemistry, material science, and related fields. An important sub-category is given by thermal simulations that describe quantum systems at finite temperatures. For example, warm dense matter (WDM)~\cite{wdm_book}, an extreme state combining high temperatures ($T\sim10^4-10^8\,$K), densities ($n\sim10^{22}-10^{27}\,$cm$^{-3}$) and pressures ($P\sim1-10^4\,$GBar), is ubiquitous throughout our universe and occurs in a host of astrophysical objects such as giant planet interiors~\cite{Benuzzi_Mounaix_2014} and brown dwarfs~\cite{becker,saumon1}. In addition, the fuel capsule in inertial confinement fusion applications has to traverse the WDM regime~\cite{hu_ICF} in a controlled way to reach ignition~\cite{Betti2016}. Consequently, WDM constitutes a highly active topic and is routinely realized in experiments at large research facilities using different techniques~\cite{falk_wdm}. Other examples for thermal quantum many-body systems of fermions include ultracold atoms~\cite{RevModPhys.80.1215,Ceperley_PRL_1992,Godfrin2012,Dornheim_SciRep_2022,filinov2023dynamic} and electrons in quantum dots~\cite{Reimann_RMP_2002,Egger_PRB_2005,PhysRevB.96.205445,Dornheim_NJP_2022}.

From a theoretical perspective, the rigorous description of such systems constitutes a difficult challenge as it must capture the complex interplay of effects such as non-ideality, quantum degeneracy and diffraction, as well as strong thermal excitations~\cite{new_POP,wdm_book}. In this context, the \textit{ab initio} path integral Monte Carlo (PIMC) method~\cite{cep} is a promising candidate as it is, in principle, capable to deliver an exact solution to the full quantum many-body problem of interest without the need for empirical input such as the exchange--correlation functional in density functional theory~\cite{Jones_RMP_2015} or the self-energy in Green function approaches~\cite{stefanucci2013nonequilibrium,Schluenzen_2020}. Unfortunately, the PIMC simulation of fermions is afflicted with the notorious fermion sign problem~\cite{troyer}, which leads to an exponential increase in the required compute time for example upon increasing the system size $N$~\cite{dornheim_sign_problem,Dornheim_2021}.

While a complete solution of the sign problem remains unlikely, the pressing need for an accurate description of quantum Fermi systems has ignited a number of promising developments in the field of quantum Monte Carlo simulations over the last decade, e.g.~Refs.~\cite{Brown_PRL_2013,Blunt_PRB_2014,Schoof_PRL_2015,Dornheim_NJP_2015,Malone_JCP_2015,Malone_PRL_2016,dornheim_prl,groth_prl,Yilmaz_JCP_2020,Joonho_JCP_2021,Hunger_PRE_2021,Hirshberg_JCP_2020,Dornheim_JCP_2020,Chin_PRE_2015,Dornheim_POP_2017,review}.
On the one hand, these methods allow for a very accurate description of a given $N$-body system. In combination with appropriate finite-size corrections~\cite{dornheim_prl,Holzmann_PRB_2016,Dornheim_JCP_2021}, this has led to the first reliable parametrizations of the uniform electron gas (UEG) covering the entire WDM parameter space~\cite{groth_prl,review,ksdt,status}, allowing for thermal DFT calculations of real WDM systems on the level of the local density approximation~\cite{karasiev_importance,kushal, Moldabekov_2023_1286} and beyond~\cite{Karasiev_PRL_2018,kozlowski2023generalized}.
On the other hand, these tools by themselves are not capable to capture long-range phenomena~\cite{Rygg_PRL_2023} that manifest on length scales larger than the given box length $L$ which, for a given density, is determined by the number of simulated particles $N$.

This precludes, for example, the description of X-ray Thomson scattering (XRTS) experiments, a key diagnostic for WDM applications~\cite{siegfried_review,kraus_xrts,Dornheim_Nature_2022,Dornheim_review}, in a forward scattering geometry where the system is probed at a small momentum transfer $q$ and, consequently, a long wavelength $\lambda=2\pi/q$. Moreover, electrical~\cite{Roepke_PRE_2021} and thermal~\cite{Jiang2023} conductivities are necessarily defined in the optical limit of $q\to0$, which makes the simulation of large systems even more important. 

Very recently, Xiong and Xiong~\cite{Xiong_JCP_2022} have proposed to alleviate the sign problem by carrying out path integral molecular dynamics (PIMD) simulations of fictitious identical particles governed by the continuous parameter $\xi\in[-1,1]$, with $\xi=1$, $\xi=0$, and $\xi=-1$ corresponding to the physically relevant cases of Bose-, Boltzmann-, and Fermi-statistics, respectively. Subsequently, Dornheim \textit{et al.}~\cite{Dornheim_JCP_xi_2023} have implemented the same idea into PIMC and found that it works remarkably well for weakly to moderately quantum degenerate systems, including the warm dense UEG. In a nutshell, these findings imply the intriguing possibility of PIMC simulations of large Fermi systems without the exponential bottleneck due to the sign problem.

\begin{figure}[t]
    \centering
    \includegraphics[width=0.49\textwidth,keepaspectratio]{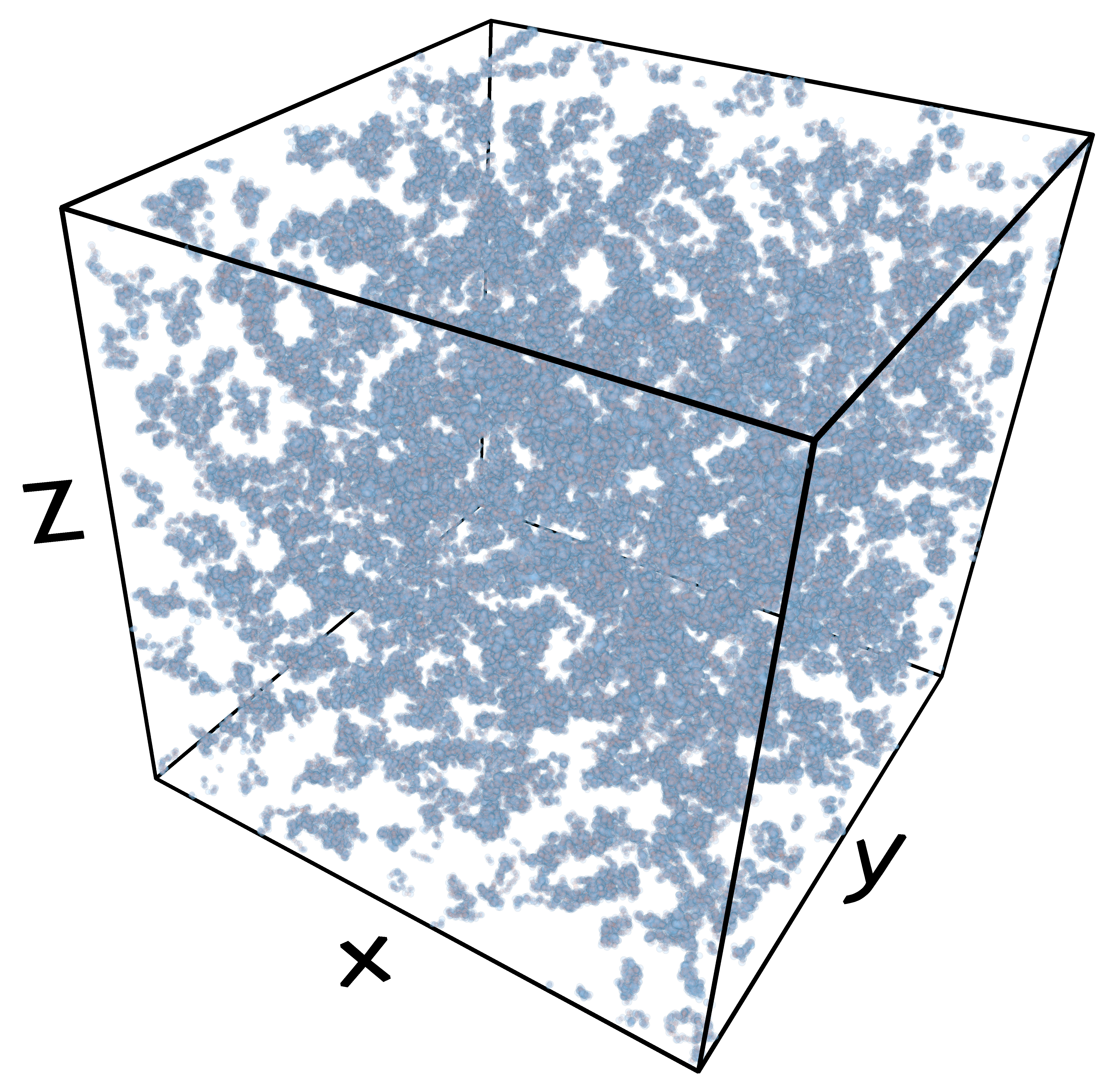}
    \caption{Snapshot from an \textit{ab initio} PIMC simulation of the warm dense UEG at $r_s=2$ and $\Theta=1$ with $N=1000$ unpolarized electrons. Each electron is represented by an entire path along the imaginary time with $P=200$ time-steps (blue spheres); their extension corresponds to the thermal wavelength $\lambda_\beta=\sqrt{2\pi\hbar^2\beta/m_e}$ and takes into account quantum effects such as diffraction.
    }
    \label{fig:snap}
\end{figure}

In the present work, we rigorously test this hypothesis by carrying out unprecedented large-scale PIMC simulations (with $N\leq10^3$, see Fig.~\ref{fig:snap}) of the UEG both at WDM parameters, and in the strongly coupled electron liquid regime~\cite{dornheim_electron_liquid,dornheim_dynamic}. In this way, we unambiguously demonstrate the capability of the $\xi$-extrapolation method to accurately describe the system on all length scales, including the difficult long-wavelength limit of $q\to0$.

These findings open up a gamut of new possibilities to study WDM, ultracold atoms, and a host of other Fermi systems.

\textbf{Path integral Monte Carlo and the $\xi$-extrapolation method.} Let us consider a system of $N=N^\uparrow + N^\downarrow$ unpolarized electrons in a cubic simulation cell of volume $V=L^3$ at an inverse temperature $\beta=1/k_\textnormal{B}T$. Writing the partition function in coordinate representation then gives
\begin{eqnarray}\label{eq:Z}
Z_{N,V,\beta} = \frac{1}{N^\uparrow!N^\downarrow!} \sum_{\sigma_{N^\uparrow}\in S_{N^\uparrow}}\sum_{\sigma_{N^\downarrow}\in S_{N^\downarrow}} \xi^{N_\textnormal{pp}} \int_V \textnormal{d}\mathbf{R}\ \bra{\mathbf{R}} e^{-\beta\hat{H}} \ket{\hat{\pi}_{\sigma_{N^\uparrow}}\hat{\pi}_{\sigma_{N^\downarrow}}\mathbf{R}}\ ,
\end{eqnarray}
where the meta-variable $\mathbf{R}=(\mathbf{r}_1,\dots,\mathbf{r}_N)^T$ contains the coordinates of both spin-up and spin-down electrons. Due to the indistinguishable nature of electrons of equal spin-orientation, we have to sum over all possible permutations of particle coordinates that are realized by the corresponding permutation operators $\hat{\pi}_{\sigma_{N^i}}$ with $i\in\{\uparrow,\downarrow\}$. A special role is played by the aforementioned continuous spin-variable $\xi$, which effectively controls the likelihood of pair exchanges (with $N_\textnormal{pp}$ being the particular number of pair exchanges needed to realize a particular permutation) in the PIMC simulation~\cite{Dornheim_permutation_cycles,Dornheim_JCP_xi_2023}. For $\xi\geq0$, all contributions to $Z$ are positive, and no sign problem occurs. In contrast, positive and negative contributions increasingly cancel for $\xi<0$, which is most severe in the fermionic limit of $\xi=-1$. 
To avoid the associated exponential decrease in the accuracy, Xiong and Xiong~\cite{Xiong_JCP_2022} have proposed to extrapolate to the latter by fitting a quadratic polynomial of the form
\begin{eqnarray}\label{eq:fit}
    O(\xi) = a_O + b_O\xi + c_O\xi^2 \ 
\end{eqnarray}
to PIMC results for an observable $\hat{O}$ in the sign-problem free domain, i.e., $\xi\geq0$. Subsequently, Dornheim \textit{et al.}~\cite{Dornheim_JCP_xi_2023} have found that this approach works remarkably well for a host of observables, including the energy $E$, static structure factor $S(q)$, and even the imaginary-time density--density correlation function $F(q,\tau)$~\cite{Dornheim_insight_2022,Dornheim_review}.

Additional details regarding the derivation of the imaginary-time PIMC method~\cite{cep} as well as the fermion sign problem~\cite{dornheim_sign_problem,Dornheim_2021} have been presented in the literature and need not be repeated here. For completeness, we note that we use a canonical adaption of the worm algorithm by Boninsegni \textit{et al.}~\cite{boninsegni1,boninsegni2} based on the extended ensemble idea presented in Ref.~\cite{Dornheim_PRB_nk_2021}. All results have been obtained for $P=200$ imaginary-time propagators with a primitive factorization scheme, and the convergence with $P$ has been carefully checked.

\begin{figure*}[t]
    \centering
    \includegraphics[width=0.49\textwidth,keepaspectratio]{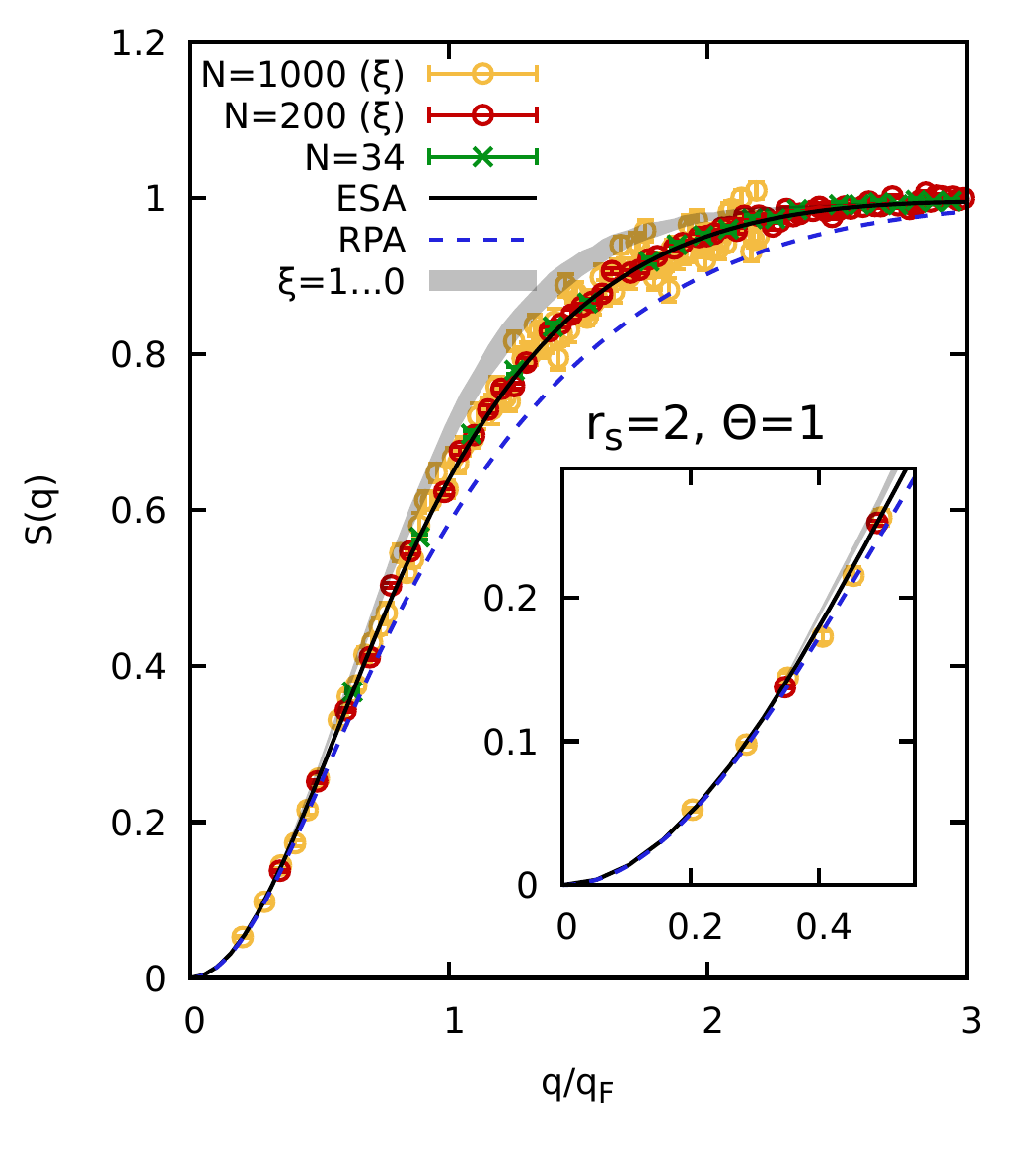}\includegraphics[width=0.49\textwidth,keepaspectratio]{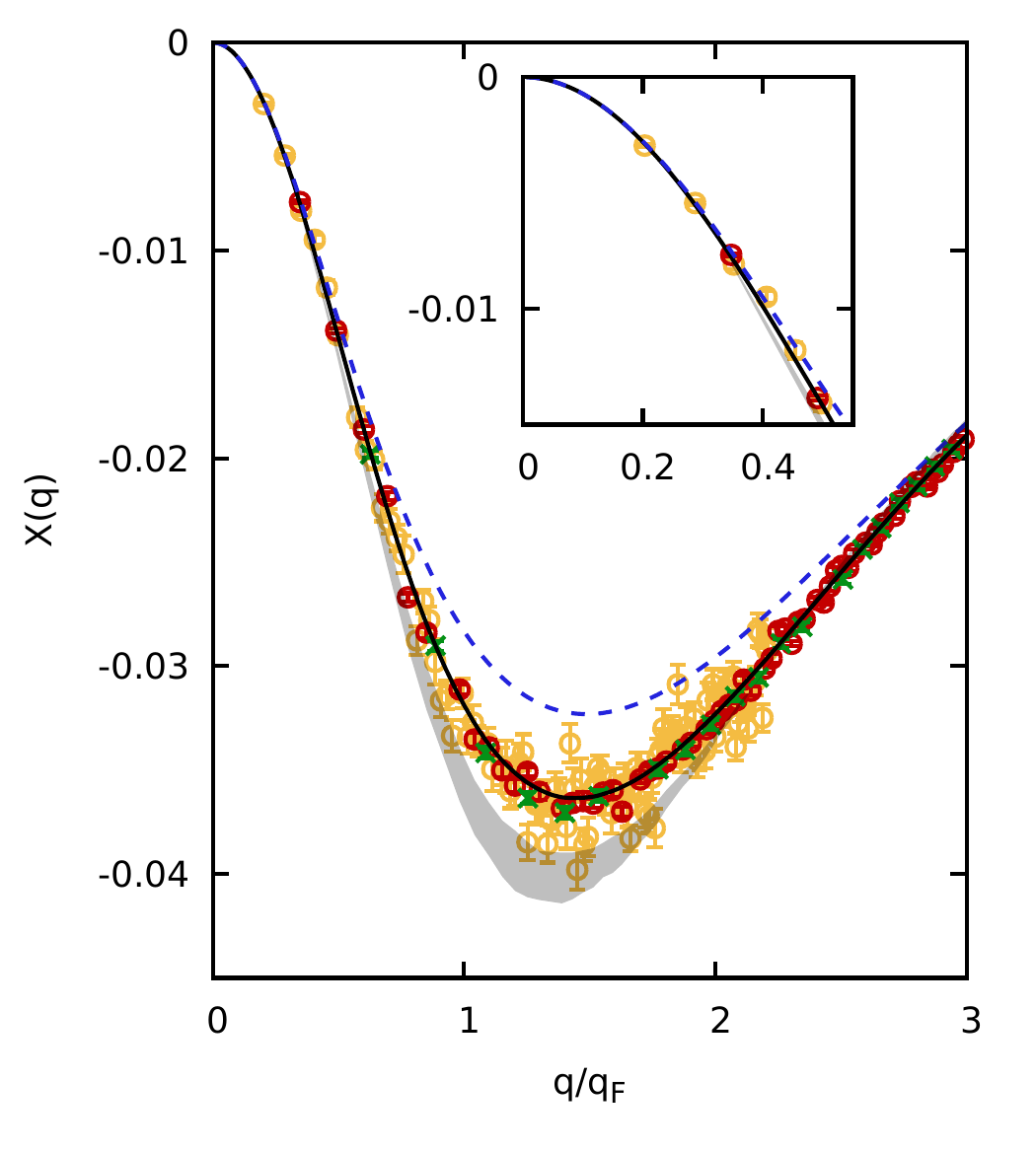}
    \caption{PIMC simulation of the warm dense UEG at $r_s=2$ and $\Theta=1$.
    Left: Static structure factor $S(q)$; right: static linear density response function $\chi(q)$ [see Eq.~(\ref{eq:chi_static})]. Dashed blue: random phase approximation (RPA); solid black: semi-empirical \textit{effective static approximation} (ESA)~\cite{Dornheim_PRL_2020_ESA,Dornheim_PRB_ESA_2021}; green crosses: exact PIMC results for $N=34$; red circles (yellow stars): new $\xi$-extrapolated PIMC results for $N=200$ ($N=1000$). The insets show magnified segments around the long-wavelength limit. 
    }
    \label{fig:UEG_rs2_theta1}
\end{figure*}

\textbf{Results.} Let us start our investigation by considering the UEG at a Wigner-Seitz radius of $r_s=2$ and degeneracy temperature $\Theta=k_\textnormal{B}T/E_\textnormal{F}=1$ (where $E_\textnormal{F}$ is the usual Fermi energy~\cite{quantum_theory}). This corresponds to a metallic density that can be realized for example in experiments with aluminium~\cite{Ramakrishna_PRB_2021} at the electronic Fermi temperature of $T=12.53\,$eV. 
In the left panel of Fig.~\ref{fig:UEG_rs2_theta1}, we show our new results for the static structure factor $S(q)$. More specifically, the dashed blue curve corresponds to the well-known random phase approximation (RPA), where the electronic density response to an external perturbation is treated on a mean-field level~\cite{quantum_theory,Dornheim_review}. For the case of the UEG, the RPA becomes exact in the long-wavelength limit of $q\to0$, but systematically underestimates the true SSF around the Fermi wavenumber $q\sim q_\textnormal{F}$. These shortcomings have been corrected in the semi-empirical \textit{effective static approximation} (ESA)~\cite{Dornheim_PRL_2020_ESA,Dornheim_PRB_ESA_2021}, which is shown as the solid black line, and which is expected to give a highly accurate description of the UEG over the entire $q$-range at these conditions.
This is indeed verified by the green crosses depicting exact PIMC results for $S(q)$ that have been obtained based on simulations with $N=34$ electrons without the $\xi$-extrapolation method. Alas, these data are only available above a minimum wavenumber of $q_\textnormal{min}=0.63q_\textnormal{F}$ which is defined by the size of the simulation cell~\cite{dornheim_prl}.

To overcome this limitation, we have carried out extensive new calculations with $N=200$ and $N=1000$ electrons using the method explored in Refs.~\cite{Xiong_JCP_2022,Dornheim_JCP_xi_2023}. In Fig.~\ref{fig:snap}, we show a snapshot from a corresponding PIMC calculation with $N=1000$. A particular strength of the method is the full treatment of quantum diffraction effects over the entire length scale, as each electron is represented by an entire path of $P=200$ positions along the imaginary-time domain; these paths would collapse to point particles in the classical limit of $\beta\to0$.

Coming back to the SSF shown in Fig.~\ref{fig:UEG_rs2_theta1}, we find that the new results for $N=200$ (red circles) and $N=1000$ (yellow circles) that have been obtained based on Eq.~(\ref{eq:fit}) using input data from the sign-problem free domain of $\xi\geq0$ (shaded grey area) are in excellent agreement with the ESA results and the exact PIMC results for $N=34$, where they are available. In particular, we unambiguously demonstrate that this approach is capable to recover the $q\to0$ limit, which is known exactly in the case of the UEG; see also the inset showing a magnified segment around this regime. The somewhats larger spread of the data points for the $N=200$ and $N=1000$ cases is due to the computational cost increasing with $\sim N^2$ and thus the significantly reduced number of snapshots available for averaging with manageable computational cost.

In the right panel of Fig.~\ref{fig:UEG_rs2_theta1}, we repeat this analysis for the static linear density response function $\chi(q)$, which can be computed from PIMC results for the ITCF $F(q,\tau)$ based on the imaginary-time version of the well-known fluctuation--dissipation theorem~\cite{bowen2,Dornheim_insight_2022}
\begin{eqnarray}\label{eq:chi_static}
    \chi(q) = -\frac{N}{V}\int_0^\beta \textnormal{d}\tau\ F(q,\tau)\quad .
\end{eqnarray}
Overall, we find the same general trends as for $S(q)$: the RPA underestimates the true magnitude of the density response around the Fermi wavenumber; the ESA is quasi-exact for all $q$ and nicely agrees with the direct PIMC results; most importantly, our new, large-scale PIMC simulations are capable to predict the correct $q\to0$ limit, see also the inset.

\begin{figure}[t]
    \centering
    \includegraphics[width=0.49\textwidth,keepaspectratio]{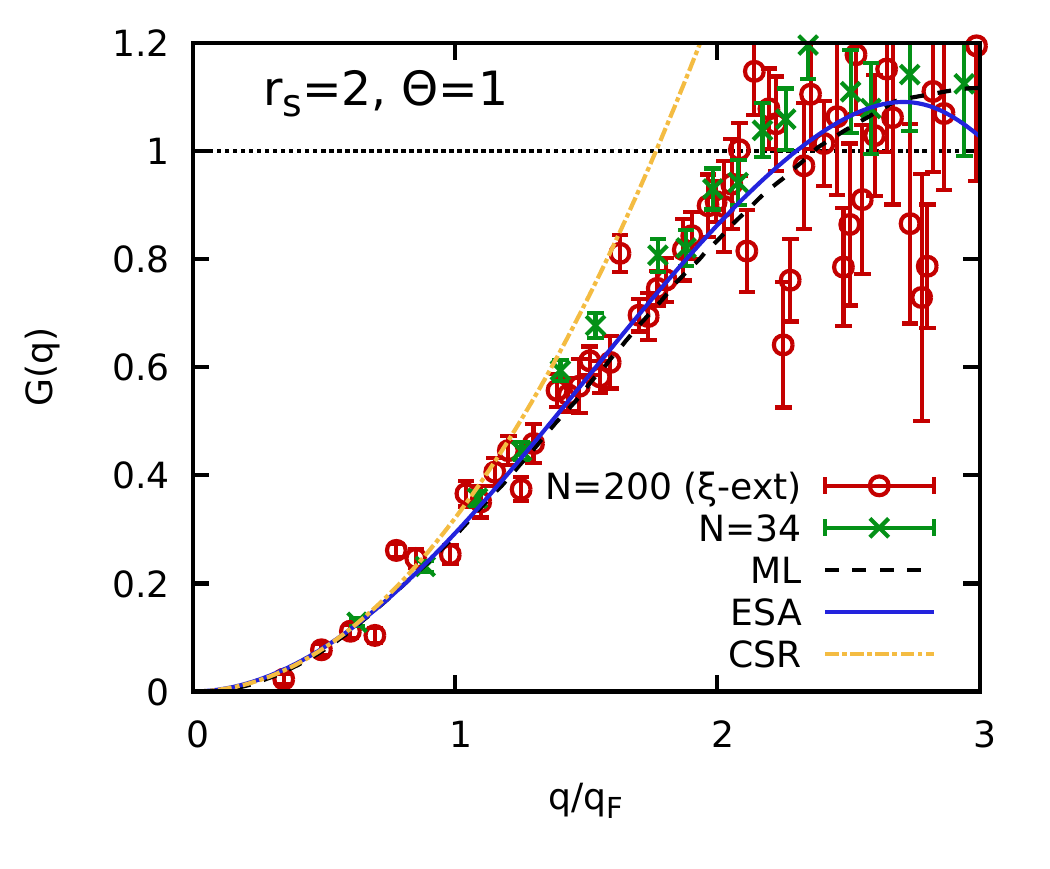}
    \caption{Local field correction $G(q)$ [see Eq.~(\ref{eq:LFC})] for $r_s=2$ and $\Theta=1$.
    Dashed black: neural-net representation from Ref.~\cite{dornheim_ML}; solid blue: analytical parametrization of the ESA~\cite{Dornheim_PRB_ESA_2021}; dash-dotted yellow: compressibility sum-rule [Eq.~(\ref{eq:CSR})] evaluated from the parametrization by Groth \textit{et al.}~\cite{groth_prl}; green crosses: exact PIMC results for $N=34$; red circles: new $\xi$-extrapolated PIMC results for $N=200$.
    }
    \label{fig:LFC}
\end{figure}

To further highlight the high quality of these results, we consider the exact expression
\begin{eqnarray}\label{eq:LFC}
    \chi(q) = \frac{\chi_0(q)}{1-\frac{4\pi}{q^2}\left[1-G(q)\right]\chi_0(q)}\quad ,
\end{eqnarray}
where $\chi_0(q)$ is the temperature-dependent Lindhard function describing the density response of an ideal Fermi gas~\cite{quantum_theory} and the complete, wavenumber resolved, information about electronic exchange--correlation effects is included in the static local field correction (LFC) $G(q)$~\cite{IIT,dornheim_ML}. Therefore, the LFC constitutes key input for a number of applications, such as the computation of electrical conductivities~\cite{Veysman_PRE_2016}, the interpretation of XRTS experiments~\cite{Fortmann_PRE_2010}, the development of thermal XC functionals \cite{Moldabekov_JPCL, Moldabekov_JCP_hybrid, JCP_2021_relevance, pribram}, and as the exchange--correlation kernel in time-dependent DFT simulations~\cite{Moldabekov_PRR_2023, Moldabekov_JCP_2023_tddft}.
In Fig.~\ref{fig:LFC}, we show the LFC for the same conditions as in Fig.~\ref{fig:UEG_rs2_theta1}. The solid blue line corresponds to the recent analytical parametrization of the ESA~\cite{Dornheim_PRB_ESA_2021} that includes the correct $q\to0$ limit given by the compressibility sum-rule (CSR), 
\begin{eqnarray}\label{eq:CSR}
    \lim_{q\to0} G(q) = -\frac{q^2}{4\pi} \frac{\partial^2}{\partial n^2}\left(n f_\textnormal{xc}\right)\ ,
\end{eqnarray}
where $f_\textnormal{xc}$ is the exchange--correlation energy of the UEG~\cite{groth_prl,review,ksdt,status}. In practice, we evaluate Eq.~(\ref{eq:CSR}) using the parametrization by Groth \textit{et al.}~\cite{groth_prl}, and the results are included as the dash-dotted yellow curve.
Furthermore, the dashed black line shows the PIMC based neural net representation of $G(q;r_s,\Theta)$ from Ref.~\cite{dornheim_ML}. The ESA and ML curves are very similar in the depicted $q$-range and only noticeably differ in the single-particle limit of $q\gg q_\textnormal{F}$. As usual, the green crosses show exact PIMC results for $N=34$ that have been obtained by inverting Eq.~(\ref{eq:LFC}), and the red circles the corresponding $\xi$-extrapolated PIMC results for $N=200$. Again, we find that the $\xi$-extrapolation method very accurately describes this sophisticated exchange--correlation property over all length scales. This raises the intriguing possibility to study long-range correlations in other Fermi systems such as warm dense hydrogen, where the true $q\to0$ limit is a subject of active investigations~\cite{Roepke_PRE_2021,Rygg_PRL_2023,Hentschel_POP_2023}.


\begin{figure*}[t]
    \centering
    \includegraphics[width=0.49\textwidth,keepaspectratio]{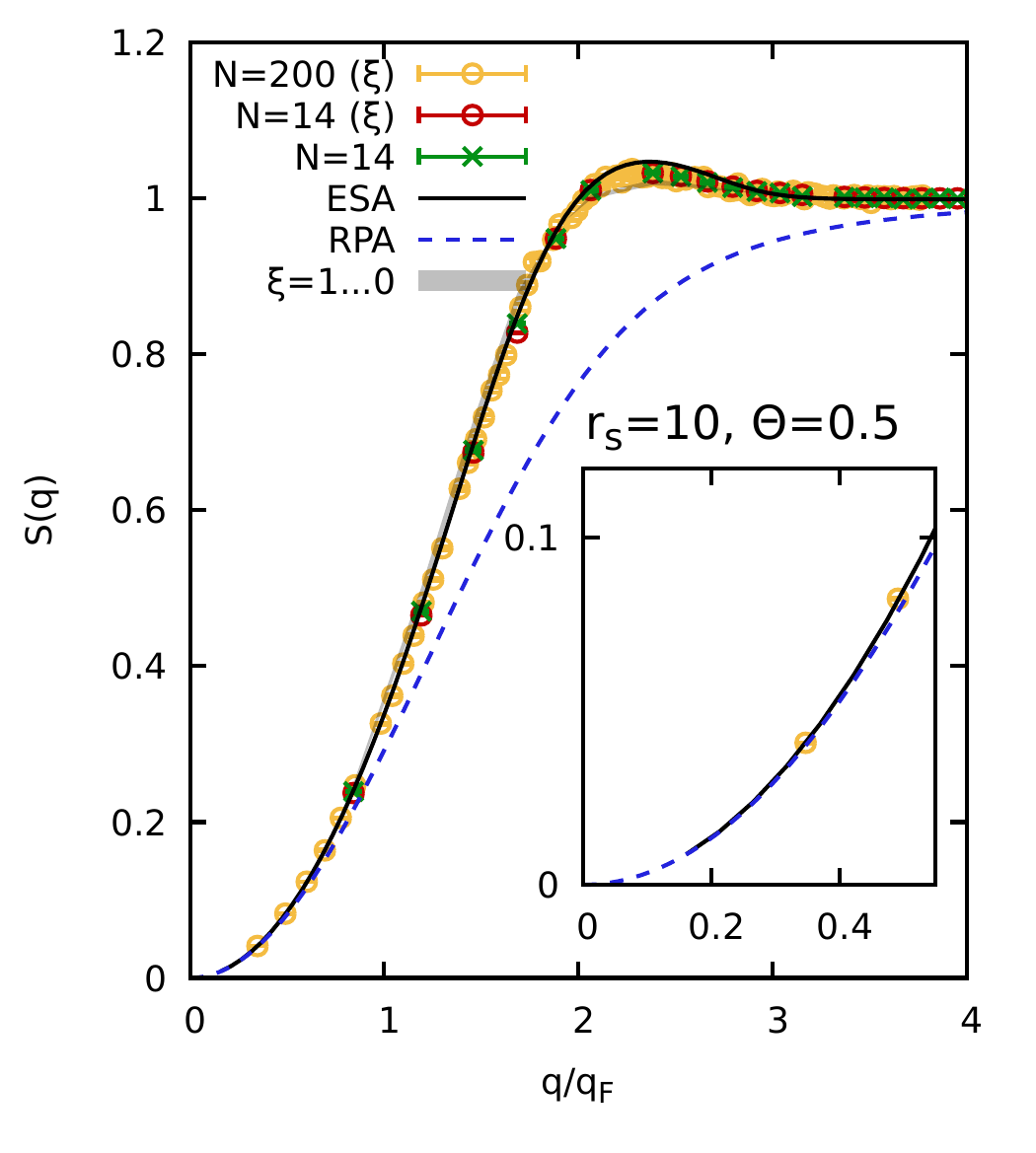} \includegraphics[width=0.49\textwidth,keepaspectratio]{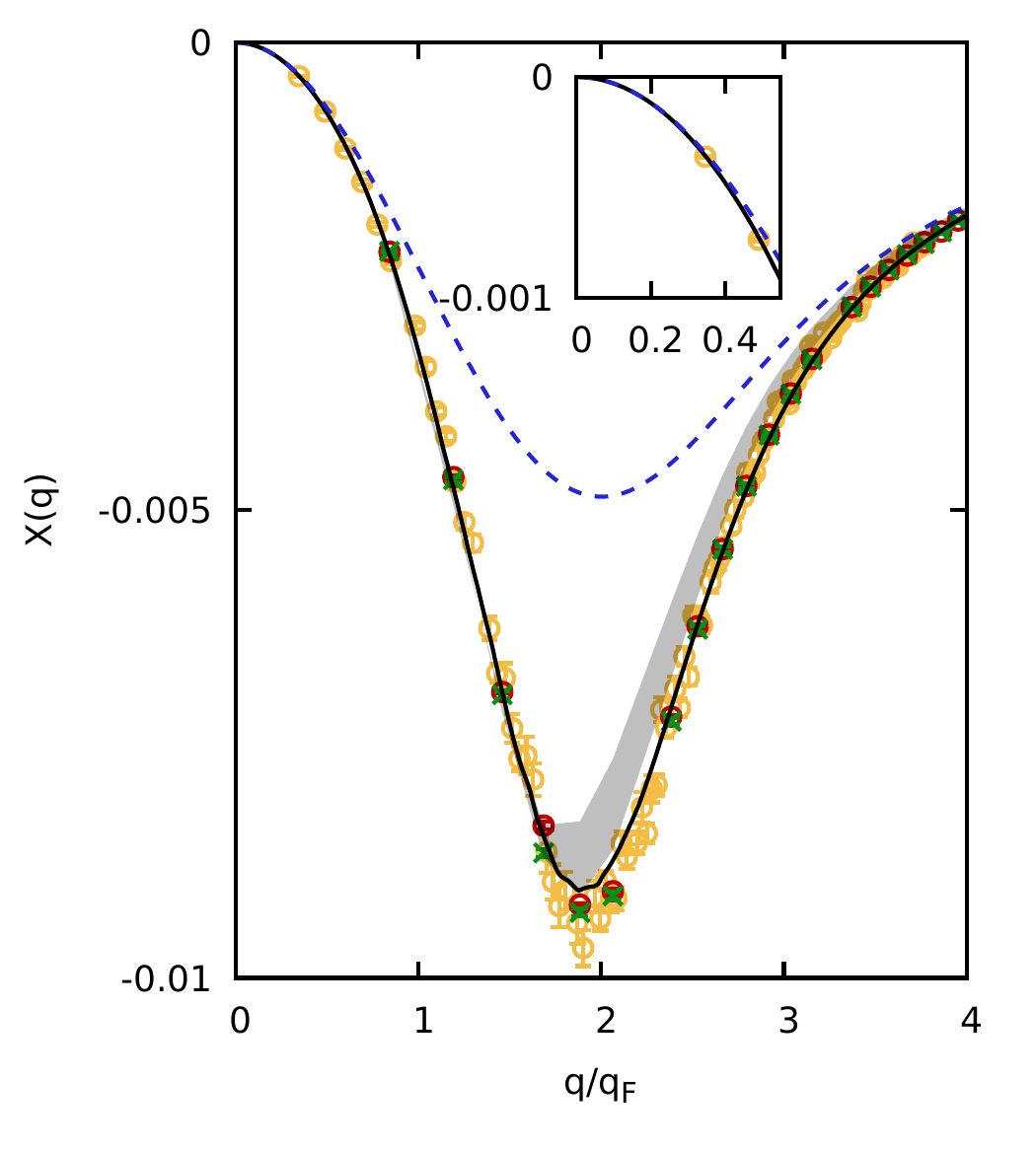}
    \caption{PIMC simulation of the warm dense UEG at $r_s=10$ and $\Theta=0.5$.
    Left: Static structure factor $S(q)$; right: static linear density response function $\chi(q)$ [see Eq.~(\ref{eq:chi_static})]. Dashed blue: random phase approximation (RPA); solid black: semi-empirical \textit{effective static approximation} (ESA)~\cite{Dornheim_PRL_2020_ESA,Dornheim_PRB_ESA_2021}; green crosses: exact PIMC results for $N=14$; red and yellow circles: $\xi$-extrapolated PIMC results for $N=14$ and $N=200$. The insets show magnified segments around the long-wavelength limit.
    }
    \label{fig:UEG_rs10_theta0p5}
\end{figure*}

As a second example, we consider the UEG at $r_s=10$ and $\Theta=0.5$. These conditions correspond to the boundary of the strongly coupled electron liquid regime~\cite{dornheim_electron_liquid,dornheim_dynamic} that is known to exhibit a wealth of interesting physical phenomena such as a non-monotonic roton-type feature in the dynamic structure factor~\cite{Takada_PRB_2016,dornheim_prl,Dornheim_Nature_2022,Dornheim_Force_2022,Hamann_PRR_2023}.
In Fig.~\ref{fig:UEG_rs10_theta0p5}, we show the corresponding SSF and static linear response function with the usual color code. To assess the accuracy of the $\xi$-extrapolation at these conditions, we have carried out new, exact PIMC simulations for $N=14$ electrons (green crosses). We find an average sign of $S=0.02$, making them computationally involved, but still feasible~\cite{dornheim_sign_problem}. 
Considering the static structure factor, the PIMC reference results are in very good agreement with the ESA curve; small systematic deviations are only visible in the vicinity of the peak where the latter slightly overestimates the true SSF. The red circles show the $\xi$-extrapolated results, which nicely agree with the green crosses everywhere. Interestingly, we find that the effect of quantum statistics on $S(q)$ is small (see the shaded grey area indicating the sign-problem free domain of $\xi\geq0$), even though the degree of cancellations of positive and negative terms is substantial already for the relatively small system size of $N=14$. A similar observation has been reported recently for the case of ultracold $^3$He~\cite{Dornheim_SciRep_2022}, which, too, constitutes a strongly coupled quantum liquid. Finally, the yellow circles show new, $\xi$-extrapolated results for $N=200$ electrons, and we find the same good performance as for the case of $r_s=2$ investigated above.

\begin{figure}[t]
    \centering
    \includegraphics[width=0.49\textwidth,keepaspectratio]{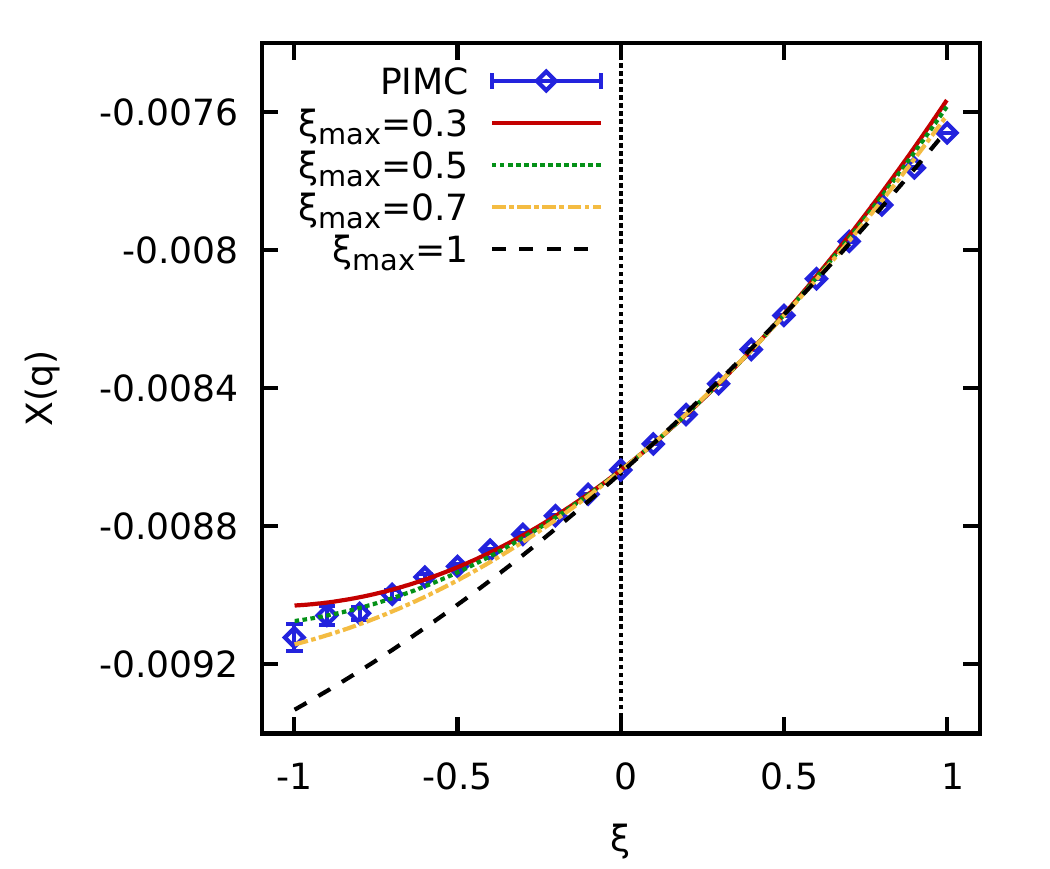} 
    \caption{$\xi$-dependence of the static linear density response function $\chi(q)$ for $N=14$, $r_s=10$, and $\Theta=0.5$ at $q\approx2q_\textnormal{F}$. The blue diamonds depict the exact PIMC results and the various curves have been obtained by quadratic fits based on Eq.~(\ref{eq:fit}) using as input data in the interval of $\xi\in[0,\xi_\textnormal{max}]$.
    }
    \label{fig:xi_range}
\end{figure}

In the right panel of Fig.~\ref{fig:UEG_rs10_theta0p5}, we show results for the static linear density response function $\chi(q)$. First, we observe that the impact of quantum statistics is considerably larger than for $S(q)=F(q,0)$. Indeed, Eq.~(\ref{eq:chi_static}) implies that $\chi(q)$ directly depends on the imaginary-time diffusion of the electrons, which is more sensitive to quantum effects compared to the static structure of the system. Second, we find that the $\xi$-extrapolation based on Eq.~(\ref{eq:fit}) becomes somewhat inaccurate when the full sign-problem free interval of $\xi\in[0,1]$ is used. This is investigated in more detail in Fig.~\ref{fig:xi_range}, where we show the $\xi$-dependence of $\chi(q)$ at $q\approx2q_\textnormal{F}$. Specifically, the blue diamonds show exact PIMC results, which are available as a benchmark over the full $\xi$-range in this case, and the various curves correspond to quadratic extrapolations based on Eq.~(\ref{eq:fit}) using as input data within different intervals $\xi\in[0,\xi_\textnormal{max}]$. The dashed black line corresponds to the usual choice of $\xi_\textnormal{max}=1$ and underestimates the true density response in the fermionic limit of $\xi=-1$ by about $1\%$. We note that this is still a reasonable level of accuracy for the description of a strongly coupled Fermi liquid, but it is considerably worse compared to the WDM case investigated above. The dash-dotted yellow and dotted green curves correspond to $\xi_\textnormal{max}=0.7$ and $\xi_\textnormal{max}=0.5$, respectively, and exhibit a substantially improved agreement with the PIMC data for $\xi<0$. Empirically, these values of $\xi$ seem equally reasonable and we have used $\xi=0.5$ to compute the red circles in the right panel of Fig.~\ref{fig:UEG_rs10_theta0p5}. Evidently, this truncated extrapolation scheme works very well over the entire $q$-range. Finally, the solid red curve in Fig.~\ref{fig:xi_range} corresponds to $\xi_\textnormal{max}=0.3$. In this case, the $\xi$-extrapolation is based on only four data points, which leads to a larger uncertainty compared to the other analyzed fitting intervals.

Returning to the static linear density response function shown in Fig.~\ref{fig:UEG_rs10_theta0p5}, we find that the effect of quantum statistics is restricted to the range of $1.75q_\textnormal{F}\lesssim q \lesssim 4q_\textnormal{F}$ for these conditions. Unsurprisingly, the $\xi$-extrapolated results for $N=200$ electrons (yellow circles) are in excellent agreement with the solid black ESA curve everywhere, and reproduce the correct $q\to0$ limit, see also the inset.
We thus conclude that the $\xi$-extrapolation method is applicable beyond the WDM regime, and constitutes a valuable tool for the investigation of strongly coupled Fermi liquids.

\textbf{Summary and outlook.} In this work, we have presented extensive new \textit{ab initio} PIMC simulations of the UEG with unprecedented system size ($N\leq1000$ electrons). In this way, we have unambiguously demonstrated that the $\xi$-extrapolation method~\cite{Xiong_JCP_2022,Dornheim_JCP_xi_2023}
is capable to describe non-ideal Fermi systems over all length scales, including the long-range limit ($q\to0$) that is known exactly in the case of the UEG, but not for other systems such as warm dense two-component plasmas or ultracold atoms. From a physical perspective, we have considered both the WDM regime that is of high current interest e.g.~for the description of laser fusion applications and astrophysical models, and the strongly coupled electron liquid regime, which exhibits phenomena such as the roton-type feature of the dynamic structure factor that is interesting in itself. We are convinced that these findings open up a host of new avenues for impactful future research in a great variety of research fields. 

First, an intriguing application of the $\xi$-extrapolation method stems from its unique capability to accurately probe even the largest length scales of the UEG. Naturally, when approaching the long-wavelength limit, otherwise inaccessible information is unlocked that concerns specific thermodynamic quantities as well as transport coefficients. (a) The $q\to0$ limit of the static local field correction directly leads to the isothermal compressibility via the eponymous sum rule without the need for thermodynamic integration or differentiations~\cite{quantum_theory}. Such a thermodynamic route that is independent from the internal energy would provide a rigorous check for existing UEG equations of state~\cite{groth_prl,status} which are typically free energy representations whose derived thermodynamic properties are known to suffer from certain pathologies~\cite{Karasiev_PRL_2018}. (b) The frequency dependent electrical conductivity is connected to the long-wavelength limit of the current response function, while the thermal conductivity is connected to the hydrodynamic limit of the heat (or energy) current response function, see the respective Kubo and Green-Kubo formulas~\cite{Kadanoff1963,Zwanzig1965}. (c) Within the current density version of linear response theory, it is known that the generalized visco-elastic coefficients are connected to long-wavelength limits that involve the respective longitudinal and transverse dynamic local field corrections~\cite{Vignale1997,Conti1999}. Moreover, the shear viscosity is connected to a hydrodynamic limit that involves the transverse local field correction~\cite{quantum_theory,Vignale1997,Conti1999}. (d) Within the spin resolved density version of linear response theory, it is known that the static spin susceptibility is connected to the $q\to0$ limit of the static spin-antisymmetric local field correction~\cite{SingwiTosi_Review}. In addition, the spin diffusion coefficient/constant is connected to a hydrodynamic limit that involves the dynamic spin-antisymmetric local field correction~\cite{Kadanoff1963}.

More important, the $\xi$-extrapolation method will allow for the first time to study long-range phenomena in real WDM systems based on \textit{ab initio} PIMC simulations starting with hydrogen~\cite{Bohme_PRL_2022,Bohme_PRE_2023,Militzer_PRE_2001,moldabekov2023bound,filinov2023equation}. This will facilitate the interpretation of XRTS experiments in a forward scattering geometry where the system is effectively probed on large length scales. From a physical perspective, one can expect that such measurements will be highly sensitive to the density of the probed system, and, in this way, will complement backward scattering measurements that are particularly sensitive to parameters such as the temperature and ionization~\cite{Tilo_Nature_2023,Dornheim_T_2022,boehme2023evidence}. In accordance with the above reasoning, a second potential application of the $\xi$-extrapolation method that is related to the study of WDM concerns the estimation of thermodynamic quantities and transport coefficients such as the thermal conductivity~\cite{Jiang2023} and the electrical conductivity~\cite{Roepke_PRE_2021}, which may substantially depend on exchange--correlation effects~\cite{Rygg_PRL_2023}.

Returning to the UEG itself, it might be interesting to study the spin-resolved density response, as well as the spin-resolved components of the static structure factor which, in contrast to the spin-averaged $\chi(q)$ and $S(q)$, are not described properly by the RPA or by more sophisticated dielectric schemes~\cite{stls,stls2,tanaka_hnc,Tanaka_CPP_2017,Tolias_JCP_2021,castello2021classical,Tolias_JCP_2023} even in the long-range limit of $q\to0$~\cite{Dornheim_PRR_2022}. Such a study might give new insights into the performance of different approximations to the LFC, and can provide useful input and benchmark data for other applications.

A further interesting line of research is given by the application of the $\xi$-extrapolation method to other systems. Based on our encouraging results for the strongly coupled electron liquid, we propose to utilize the method to ultracold atoms such as the short-range interacting $^3$He, but also dipolar systems~\cite{Dornheim_PRA_2020,dornheim_sign_problem}. For these systems, interesting long-range phenomena include the acoustic mode in the dynamic structure factor for small $q$, and the long-wavelength limit of the SSF that is given by the compressibility sum-rule~\cite{Ferre_PRB_2016}. Furthermore, we note that the method has already been applied to electrons in quantum dots in previous studies~\cite{Xiong_JCP_2022,Dornheim_JCP_xi_2023,Xiong_PRE_2023}. Indeed, it is well known that trapped Fermi systems exhibit interesting effects such as the formation of Wigner molecules~\cite{Egger_PRL_1999,PhysRevB.83.085409} and a negative superfluid fraction~\cite{Blume_PRL_2014,Dornheim_PRA_2020,Dornheim_NJP_2022}.

Finally, we stress that, despite its impressive performance, the $\xi$-extrapolation method is a very new technique, and its further methodological improvement remains in its infancy compared to more established methods such as restricted PIMC~\cite{Ceperley1991,Brown_PRL_2013,Driver_PRL_2015}. The observed improvement of the extrapolation due to the truncation of the $\xi$-interval in the sign-problem free domain for $r_s=10$ and $\Theta=0.5$ suggests that much can potentially be gained by developing better extrapolation schemes. In addition, the absence of the exponential bottleneck poses new challenges, which might be met by combining PIMC with other ideas such as the adaptive long-range potential scheme that has been suggested in the recent Ref.~\cite{Muller_PRX_2023}, or the quantum ring-polymer contraction method that has been introduced in the context of PIMD~\cite{John_PRE_2016}.

\section*{Acknowledgments}
This work was partially supported by the Center for Advanced Systems Understanding (CASUS) which is financed by Germany’s Federal Ministry of Education and Research (BMBF) and by the Saxon state government out of the State budget approved by the Saxon State Parliament. 
This work has received funding from the European Research Council (ERC) under the European Union’s Horizon 2022 research and innovation programme
(Grant agreement No. 101076233, "PREXTREME").
The PIMC calculations were partly carried out at the Norddeutscher Verbund f\"ur Hoch- und H\"ochstleistungsrechnen (HLRN) under grant mvp00024, on a Bull Cluster at the Center for Information Services and High Performance Computing (ZIH) at Technische Universit\"at Dresden, and on the
HoreKa supercomputer funded by the Ministry of Science, Research and the Arts Baden-W\"urttemberg and
by the Federal Ministry of Education and Research.

\bibliography{ref}

\end{document}